\documentclass[12pt]{article}
\usepackage{a4wide}
\usepackage{epsfig}

\newcommand{\eps}{{\epsilon}} 

\newcommand{\cH}{\mathcal{H}} %
\newcommand{\ep}{{\epsilon}} 
\newcommand{\MS}{{\mathrm{MS}}}

\def\z#1{{\zeta_{#1}}}
\def\ca{{C^{}_A}}
\def\cf{{C^{}_F}}
\def\nf{{n^{}_{\! f}}}

\def\T2F{{T^{\,2}_{\! F}}}
\def\M{N}
\newcommand{\bra}[1]{\langle#1\mid}
\newcommand{\ket}[1]{\mid#1\rangle}

\begin{document}
\thispagestyle{empty}

\begin{center}
 {\Large\textbf{The 16th moment of the three loop \\ anomalous dimension of the non--singlet \\ transversity operator in QCD}

 }\vspace{15mm}

{\sc A.~A.~Bagaev$^a$, A.~V.~Bednyakov$^b$, A.~F.~Pikelner$^b$ and V.~N.~Velizhanin$^c$}\\[5mm]

\emph{$^a$Saint Petersburg State University\\ 199034 St.~Petersburg, Russia}\\[5mm]

\emph{$^b$Joint Institute for Nuclear Research\\ 141980 Dubna, Russia}\\[5mm]

\emph{$^c$Theoretical Physics Department, Petersburg Nuclear Physics Institute\\ Orlova Roscha, Gatchina, 188300 St.~Petersburg, Russia}\\[5mm]

\textbf{Abstract}\\[2mm]
\end{center}
 \noindent
We present the result of the three loop anomalous dimension of non--singlet transversity operator in QCD for the Mellin moment $N=16$.
The obtained result coincides with the prediction from \texttt{arXiv:1203.1022} and can serve as a confirmation of the correctness of the general expression for three loop anomalous dimension of non--singlet transversity operator in QCD for the arbitrary Mellin moment.
 \newpage

\setcounter{page}{1}
A general expression for the three loop anomalous dimension of non--singlet transverse operator in QCD for the arbitrary Mellin moment was obtained
recently~\cite{Velizhanin:2012nm}. It has been reconstructed from the known first fifteen moments with the help of LLL-algorithm~\cite{Lenstra:1982}.
In spite of the fact that the reconstructed result seems to be correct it would be nice
to have an additional confirmation from the direct calculations of higher moments, starting from $N=16$.

The local flavor non--singlet twist-2 transversity operator is given by
  \begin{equation}\label{op2}
   {\mathcal O}_{q,r}^{{\mathrm {TR,ns}}, \mu, \mu_1, \ldots, \mu_\M}(z) =\frac{1}{2}\,i^{N-1} \hat S
   \left[\overline{q}(z) \sigma^{\mu \mu_1} {\mathcal D}^{\mu_2} \ldots {\mathcal D}^{\mu_\M} \frac{\lambda_r}{2}\, q(z)\right],
  \end{equation}
where $\sigma^{\mu\nu} = (i/2)\left[\gamma^\mu \gamma^\nu - \gamma^\nu \gamma^\mu \right]$, $\lambda_r$ are the Gell-Mann matrices for $SU(3)_{\rm flavor}$,
${\mathcal D}^{\mu}$ corresponds to the covariant derivative in QCD, $q (\overline{q})$ denote the quark and antiquark fields, and the operator $\hat S$ symmetrizes the Lorentz indices and subtracts the trace terms. This operator appears, for example, in the study of semi-inclusive deeply inelastic scattering (SIDIS) \cite{Ralston:1979ys, Jaffe:1991kp, Jaffe:1991ra, Cortes:1991ja} and in the polarized Drell---Yan process \cite{Cortes:1991ja, Collins:1992kk, Jaffe:1993xb, Tangerman:1994bb, Boer:1997nt, Artru:1989zv}. The scaling violation of the transversity distribution was explored in leading \cite{Artru:1989zv, Baldracchini:1980uq, Shifman:1980dk, Bukhvostov:1985rn, Blumlein:2001ca, Mukherjee:2001zx} and next-to-leading order \cite{Hayashigaki:1997dn,Kumano:1997qp,Vogelsang:1997ak}. At the three-loop order first 15 Mellin moments were obtained in Refs.~\cite{Gracey:2003yr,Gracey:2003mr,Gracey:2006zr,Gracey:2006ah,Blumlein:2009rg,Velizhanin:2012nm}.

The calculation of the 16th moment for the anomalous dimension associated with transversity operator~(\ref{op2}) is several times more difficult than the
calculation of the corresponding moment for non--singlet anomalous dimension associated with the deeply inelastic structure functions $F_2(x,Q^2)$ and
$F_L(x,Q^2)$, which was obtained at third order in Ref.~\cite{Blumlein:2004xt} by method from Refs.~\cite{Larin:1991fx,Larin:1993vu}. To perform this kind of
computation we had to improve substantially the method used for the calculation of lower moments in Ref.~\cite{Velizhanin:2012nm}. In this paper we give a detailed description of our calculations.

The anomalous dimension of the operator in $\MS$-like scheme can be obtained from the corresponding renormalization constant by means of the following general formula
 \begin{equation}\label{defga}
   \gamma_{\mathrm {TR,ns}}
   =g^2\frac{\partial}{\partial g^2}\, c^{(1)}_{\mathrm {TR,ns}}(\alpha,g^2)=\sum^\infty_{n=1}
   \gamma_{\mathrm {TR,ns}}^{(n-1)}(\alpha,g^2)g^{2n}.
 \end{equation}
  where $g$ stands for strong gauge coupling, $\alpha$ is a gauge fixing parameter, and $c^{(1)}_{\mathrm{ TR, ns}}(\alpha,g^2)$
  is the coefficient of the lowest pole in $\eps$ in the Laurent expansion of the renormalization constant for the operator
 \begin{equation}\label{eq:5}
      Z_{\mathcal{O}^\mathrm{TR,ns}}\!\left(\frac{1}{\epsilon},\alpha,g^2\right)=1+\sum^\infty_{n=1}c_{\mathrm{ TR, ns}}^{(n)}(\alpha,g^2)\epsilon^{-n}.
 \end{equation}

For calculation of the renormalization constants, following \cite{Larin:1993tp} (see also~\hbox{\cite{Tarasov:1976ef,Vladimirov:1979zm,Tarasov:1980kx}}), we use the
multiplicative renormalizability of corresponding Green's functions. The renormalization constants $Z_\Gamma$ relate the dimensionally regularized
one-particle-irreducible Green's function with renormalized one as:
 \begin{equation}\label{multren}
   \Gamma_{\mathrm{Renormalized}}\left(\frac{Q^2}{\mu^2},\alpha,g^2\right)=\lim_{\epsilon \rightarrow 0}
   Z_\Gamma\left(\frac{1}{\epsilon},\alpha,g^2\right)
   \Gamma_{\mathrm{Bare}}\left(Q^2,\alpha_{\mathrm{B}},g^2_\mathrm{B},\epsilon\right),
 \end{equation}
where $g^2_{\mathrm{B}}$ and $\alpha_{\mathrm{B}}$ are the bare charge and the bare gauge fixing parameter, correspondingly, with
 \begin{equation}
 g^2_{\mathrm{B}}=\mu^{2\epsilon}\left[g^2+\sum_{n=1}^{\infty}a^{(n)}\!\left(g^2\right)\epsilon^{-n}\right],\qquad
 \alpha_{\mathrm{B}}=\alpha Z_3\label{gbex}
 \end{equation}
and $Z_3$ being the gluon field renormalization constant.

In order to calculate the required anomalous dimension, following~\cite{Bierenbaum:2009mv}, we consider the Green's function $\hat{G}^{ij,{\mathrm {TR,ns}}}_{\mu,q}$, which is obtained by contracting the matrix element of the local operator~(\ref{op2}) with the source term $J_N = \Delta^{\mu_1} \ldots \Delta^{\mu_\M}$
  \begin{equation}\label{GijTRNS}
   \overline{u}(p,s) G^{ij, {\mathrm {TR,ns}}}_{\mu,q} \lambda_r u(p,s)=J_\M\bra{q_i(p)}{\mathcal O}_{q,r;\mu, \mu_1, \ldots, \mu_\M}^{{\mathrm {TR,ns}}}\ket{q^j(p)}\,,
  \end{equation}
where $p$ and $s$ denote the four-vectors of the momentum and spin of the external quark line, $u(p,s)$ is the corresponding bi-spinor, $\Delta^2 = 0$. The
contraction with the source term $J_N = \Delta^{\mu_1}\ldots \Delta^{\mu_\M}$ allows to write a general expression for the corresponding projector, which can be found in Ref.~\cite{Bierenbaum:2009mv}.

The unrenormalized Green's function has the following Lorentz structure~\cite{Blumlein:2009rg}
 \begin{eqnarray}
   \hat{G}^{ij, {\mathrm {TR,ns}}}_{\mu,q}&=&\delta_{ij}(\Delta \cdot p)^{\M-1} \Big( \Delta_{\rho}\sigma^{\mu\rho} \,\Sigma^{\mathrm {TR,ns}}(p) +c_1 \Delta^\mu + c_2
   p^\mu +c_3 \gamma^\mu p \hspace*{-2mm} / \nonumber \\
   &&\hspace*{35mm}+c_4 \Delta \hspace*{-3mm}/~p \hspace*{-2mm}/ \Delta^\mu +c_5 \Delta \hspace*{-3mm}/~p\hspace*{-2mm}/ p^\mu \Big),
 \end{eqnarray}
with unphysical constants ${c_k|}_{k=1 \ldots 5}$. To determine $\Sigma^{\mathrm {TR,ns}}(p)$ we use the following projector (see~\cite{Blumlein:2009rg})
 \begin{eqnarray}
   \Sigma^{\mathrm {TR,ns}}(p)&=&
   - i \frac{\delta^{ij}}{4N_c(\Delta\cdot p)^{\M+1} (D-2)}
   \Big\{
   {\mathrm {Tr}}[ \Delta\hspace*{-3mm}/~p\hspace*{-2mm}/~
   p^{\mu}\hat{G}^{ij, {\mathrm {TR,ns}}}_{\mu,q}]
   -(\Delta\cdot p) {\mathrm {Tr}}[p^{\mu}\hat{G}^{ij, {\mathrm {TR,ns}}}_{\mu,q}] \nonumber\\
  && \hspace*{55mm}+i(\Delta\cdot p) {\mathrm {Tr}}[\sigma_{\mu \rho} p^\rho
   \hat{G}^{ij, {\mathrm {TR,ns}}}_{\mu,q}]
   \Big\}\,,\label{eqc3}
 \end{eqnarray}
where $N_c$ denotes the number of colors and $D = 4 - 2 \eps$.

Once the renormalization constant for $\Sigma^{\mathrm {TR,ns}}(p)$ is known, it is straightforward to compute $Z_{\mathcal{O}^\mathrm{TR,ns}}$  by means of
 \begin{equation}\label{gbz}
   Z_{\mathcal{O}^{\,\mathrm {TR,ns}}}=Z_{q}^{-1}\,Z_{\Sigma^{\,\mathrm {TR,ns}}},
 \end{equation}
with $Z_q$ being the renormalization constant for the quark field.

As in our previous calculations~\cite{Velizhanin:2012nm} we made use of the program \texttt{DIANA}~\cite{Tentyukov:1999is}, which calls
\texttt{QGRAF}~\cite{Nogueira:1991ex}  to generate all diagrams and the \texttt{FORM} package \texttt{COLOR}~\cite{vanRitbergen:1998pn} for evaluation of the
color traces. The unrenormalized three-loop $\Sigma^{\mathrm {TR,ns}}(p)$ was computed with the help of the \texttt{FORM}~\cite{Vermaseren:2000nd} package
\texttt{MINCER}~\cite{Gorishnii:1989gt}. Corresponding renormalization constant $Z_{\Sigma^{\mathrm {TR,ns}}}$ was determined from the requirement that the poles in $\ep$ cancel in the r.h.s. of Eq.~(\ref{multren}).

Since the right-hand side of Eq.~(\ref{multren}) contains the bare gauge fixing parameter $\alpha_{\mathrm{B}}$, we should perform all calculations up to two loops with the arbitrary gauge fixing parameter~$\alpha$ (i.\,e. the propagator of gluon is $(g_{\mu\nu}-(1-\alpha)q_\mu q_\nu/q^2)/q^2$), while for the three-loop calculations it is sufficient to use the Feynman gauge $\alpha=1$. To obtain the result for the renormalization constant from Eq.~(\ref{multren}), we put $\alpha=1$ only after expansion in the right-hand side of Eq.~(\ref{multren}).

It is obvious from the discussion presented above that our main task was to extract $\Sigma^{\mathrm {TR,ns}}(p)$ contribution from the unrenormalized Green function for the transersity operator.
A convenient way to do this is to get rid of the contraction with $\Delta$ in Eqs.~(\ref{GijTRNS}) and~(\ref{eqc3}) and perform the Passarino\hbox{---}Veltman decomposition of the resulting tensor in the following way (see~\cite{Bierenbaum:2009mv}):
 \begin{equation}\label{GenResInt}
   R_{\{\mu_1\ldots\mu_{16}\mu_{17}\}} \ \equiv\ \sum_{j=1}^{9} A_j\Bigl(\prod_{k=1}^{j-1} g_{\{\mu_{2k}\mu_{2k-1}} \Bigr)
   \Bigl(\prod_{l=2j-1}^{17} p_{\mu_l\}} \Bigr),
 \end{equation}
where $\{\}$ means the symmetrization with respect to the indices.
It should be pointed out that Eq.~(\ref{GenResInt}) has one additional index in comparison with the operator in Eq.~(\ref{op2}) as projector~(\ref{eqc3}) itself contains an additional vector $\Delta$.
Due to the fact that the contraction with $g_{\mu\nu}$ gives $\Delta^2 = 0$ it is sufficient to calculate only the coefficient $A_1$ from Eq.~(\ref{GenResInt}).
The corresponding projector reads~\cite{Bierenbaum:2009mv} (note again, that the projector has 17 indices for the 16th moment):
 \begin{eqnarray}
   \Pi_{\mu_1\ldots\mu_{17}}&=&F(17)\sum_{i=1}^{9}C(i,17)\Big(\prod_{l=1}^{9-i}\frac{g_{\mu_{2l-1}\mu_{2l}}}{p^2}\Big) \Big(\prod_{k=19-2i}^{17}\frac{p_{\mu_k}}{p^2}\Big),\label{Proj1}\\[3mm]
C(k,17) &=& (-1)^{k}
                 \frac{2^{2(k-5)}\Gamma(18)\Gamma(D/2+7+k)}
                      {\Gamma(10-k)\Gamma(2k)\Gamma(D/2+8)}\,, \\[2mm]
F(17)  &=& \frac{\Gamma(D/2+1/2)}
                        {2^7 (D-1)\Gamma(D/2+15/2)}
 \end{eqnarray}
and the general expression for the arbitrary $N$ can be found in Ref.~\cite{Bierenbaum:2009mv}.

It turns out that for the calculation of the 16th moment the above-mentioned procedure has to be substantially optimized. The main difficulties for higher moments are the following.
Firstly, the number of terms for an operator vertex with some outgoing gauge lines increases very fast (see Appendix of Ref.~\cite{Bierenbaum:2009mv}).
Secondly, the number of terms in the projectors grow even more rapidly.
However, since the projector~(\ref{Proj1}) when applied to a diagram acts only on external Lorentz indices of the operator vertex, the contraction
 \begin{equation}\label{ProjctOp}
   \Pi\textsf{O}^{\,\mu,\,{\mathrm {TR,ns}}}_{\,\nu,\,{q,r}}=\Pi_{\mu_1\ldots\mu_{17}}{\mathcal O}_{q,r}^{{\mathrm {TR,ns}}, \mu, \mu_1, \ldots, \mu_{16}}\,,
   \qquad\qquad
   \mu_{17}=\nu
 \end{equation}
is universal for all diagrams and we can substitute this expression (with proper relabeling) in place of the operator vertex.
It should be stressed that such contraction produces only several Lorentz structures for arbitrary moment $N$.
For example, in the case of an operator vertex without gauge lines attached, i.\,e.,  when all covariant derivatives in operator~(\ref{op2}) are substituted by the ordinary ones, we have 10 Lorentz structures multiplied by combinations of scalar products of internal and loop momenta
 \begin{eqnarray}
   \Pi\textsf{O}^{{\mathrm {TR,ns}}}_{\mu\nu,\,{q,r}}(0)& = & \hat Q\gamma_\mu Q_\nu \,\cH_1(p,Q) + \hat Q\gamma_\mu p_\nu \,\cH_2(p,Q) +
   \hat p\gamma_\mu Q_\nu\,\cH_3(p,Q) + \hat p\gamma_\mu p_\nu \,\cH_4(p,Q) \nonumber\\[1mm]
    & & {}+ \gamma_\mu\hat Q Q_\nu \,\cH_5(p,Q) + \gamma_\mu\hat Q p_\nu \,\cH_6(p,Q) + \gamma_\mu\hat pQ_\nu \,\cH_7(p,Q) + \gamma_\mu\hat p p_\nu\,
    \cH_8(p,Q) \nonumber\\[1mm]
    & & {}+ \gamma_\mu\gamma_\nu \,\cH_9(p,Q) + \gamma_\nu\gamma_\mu \,\cH_{10}(p,Q)\,,
 \end{eqnarray}
where $Q$ corresponds to external momentum, and $p$ stands for loop momenta flowing through the operator. In \texttt{MINCER} notations, $p=p_1,\ldots,p_8$.
For operator vertex with one outgoing gauge line there are 72 Lorentz structures with 3 different indices, for operator vertex with two gauge fields we have 710 Lorentz structures with 4 different indices and for  operator vertex  with three gauge fields the number of structures with 5 different indices is 8900 (the total number is 9692).
In general case the functions $\cH_k$ depend on scalar products of momenta entering to operator vertex. Thus, we have created a file containing the expressions for coefficients $\cH_k$ and substituted them in a given diagram only after contraction of Lorentz indices and computation of traces of $\gamma$-matrices
products.
After these kind of substitutions the output can be processed by \texttt{MINCER}. This procedure works well for all diagrams with no more than 2 gauge
lines at operator vertex, i.\,e. we calculated 668 from 682 three-loops diagrams\footnote{The results for all diagrams can be obtained from the authors upon request.}. Elapsed computing time was equal to 400 hours of CPU time.

\begin{figure}[t]
  \epsfig{file=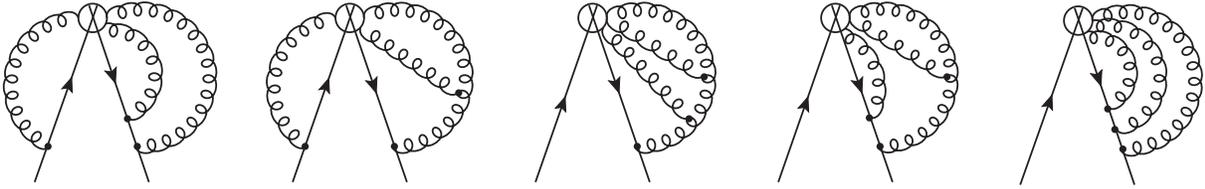,width=16cm,angle=0}
  \vspace*{4mm}
  \caption{Examples of diagrams containing three gauge fields in the operator vertex}
  \label{fig:diag3}
 \end{figure}

However, for the diagrams containing three gauge lines  in operator vertex the described procedure requires too much computer time. Fortunately, taking into account diagram symmetry there are only 7 such graphs (some of them are presented in Fig.~\ref{fig:diag3}).

Each vertex of this type includes six different structures corresponding to all permutations of gauge fields (see Appendix of Ref.~\cite{Bierenbaum:2009mv}). Taking into account the color symmetry we have 15 different Lorentz structures for all 7 diagrams, containing such operator vertex. Each of these 15 structures can be obtained from the same combination of momenta and Lorentz indices. The difference between structures consists only in momenta labeling. A template was created for a chosen momenta combination and then other structures were obtained by means of proper substitution of momenta with the help of \texttt{sed} stream editor. The obtained data was used to declare a \texttt{Table} in \texttt{FORM} as the momenta are fixed for each structure. It turned out that the usage of \texttt{Table} allowed us to speed up by several orders of magnitude the process of generation of scalar expressions to be processed by \texttt{MINCER}. However, even after such an optimization the most complicated structures, which appear in the first three diagrams presented in Fig.~\ref{fig:diag3}, required more than 30 hours of CPU time with \texttt{FORM} and more than 200~Gb HDD space (for temporary files). All diagrams containing the operator vertex with three gauge lines were calculated in approximately 250 hours.

The final expression for 16th Mellin moment of the three-loop anomalous dimension of the flavor non--singlet transversity operator~(\ref{op2}) was found to be given by
 \begin{eqnarray}
   \gamma_{\mathrm{TR,ns}}(16) & = & a_s\,\cf\frac{1896019}{180180}+a_s^2\,\Bigg(-\frac{2038896291251}{194788994400}\,\cf\nf \nonumber\\
    & & {}+\frac{97129786174481}{2181636737280}\,\ca\cf- \frac{187897192426587961}{23398054007328000}\,\cf^2\Bigg) \nonumber\\
    & & {}+a_s^3\Bigg(-\frac{9588431966492629867}{3579902263121184000}\,\cf\nf^2 \nonumber\\
    & & {}-\ca\cf\nf\left(\frac{26209369318544258455663}{1002372633673931520000}+\frac{3792038}{45045}\,\z3\right) \nonumber\\
    & & {}-\cf^2\nf\left(\frac{214170060836708111694659}{2810574247360239360000}-\frac{3792038}{45045}\,\z3\right) \nonumber\\
    & & {}-\cf^3\left(\frac{5187682484935485448445159707}{303845560733620756730880000}-\frac{100658462371}{2705402700}\,\z3\right) \nonumber\\
    & & {}-\ca\cf^2\left(\frac{24000075522445759616746817}{590220591945650265600000}+\frac{100658462371}{1803601800}\,\z3\right) \nonumber\\
    & & {}+\ca^2\cf\left(\frac{1448460151630990534353391}{5613286748574016512000}+\frac{100658462371}{5410805400}\,\z3\right)\Bigg)\,,
 \end{eqnarray}
where $a_s=\alpha_s /(4\pi) = g^2/(16\pi^2)$, $\z3 \equiv \zeta(3)$ is the Riemann zeta function, $\ca=N_c$ and $\cf=(N_c^2-1)/(2N_c)$ are quadratic Casimir operators for gauge group $SU(N_c)$, $\nf$ is the number of active quark flavors and we put $T_F=1/2$.
This expression is in a full agreement with the prediction given in Ref.~\cite{Velizhanin:2012nm}, and can serve as a direct confirmation of the general result~\cite{Velizhanin:2012nm} for arbitrary Mellin moment $N$.

\subsection*{Acknowledgments}
This work is supported by RFBR grants 10-02-01338-a, 12-02-00412-a, RSGSS-65751.2010.2.

\end{document}